# Atomistic mechanism for trapped-charge driven degradation of perovskite solar cells

Kwisung Kwak[1,2,†], Eunhak Lim[3,†], Namyoung Ahn[1,2,†], Jiyoung Heo[4], Kijoon Bang[1,2], Seong Keun Kim[3*], and Mansoo Choi[1,2*]

[1]*Global Frontier Center for Multiscale Energy Systems, Seoul National University, Seoul 08826, Korea*
[2]*Department of Mechanical and Aerospace Engineering, Seoul National University, Seoul 08826, Korea*
[3]*Department of Chemistry, College of Natural Sciences, Seoul National University, Seoul 08826, Korea*
[4]*Department of Green Chemical Engineering, Sangmyung University, Chungnam 31066, Korea*

[†]*These authors contributed equally to this work.*
*\*To whom correspondence should be addressed. E-mail: seongkim@snu.ac.kr, mchoi@snu.ac.kr*

# Abstract

It is unmistakably paradoxical that the weakest point of the photoactive organic-inorganic hybrid perovskite is its instability against light. Why and how perovskites break down under light irradiation and what happens at the atomistic level of these materials during the degradation process still remains unanswered. In this paper, we revealed the fundamental origin and mechanism for irreversible degradation of hybrid perovskite materials from our new experimental results and *ab initio* molecular dynamics (AIMD) simulations. We found that the charges generated by light irradiation and trapped along the grain boundaries of the perovskite crystal result in oxygen-induced irreversible degradation in air even in the absence of moisture. The present result, together with our previous experimental finding on the same critical role of trapped charges in the perovskite degradation under moisture, suggests that the trapped charges are the main culprit in both the oxygen- and moisture-induced degradation of perovskite materials. More detailed roles of oxygen and water molecules were investigated by tracking the atomic motions of the oxygen- or water-covered methylammonium lead triiodide ($MAPbI_3$ for $CH_3NH_3PbI_3$) perovskite crystal surface with trapped charges via AIMD simulation. In the first few picoseconds of our simulation, trapped charges start disrupting the crystal structure, leading to a close-range interaction between oxygen or water molecules and the compositional ions of $MAPbI_3$. We found that there are different degradation pathways depending both on the polarity of the trapped charge and on the kind of gas molecule. Especially, the deprotonation of organic cations was theoretically predicted for the first time in the presence of trapped anionic charges and water molecules. Additionally, we confirmed that a more structurally stable, multi-component perovskite material (with the composition of $MA_{0.6}FA_{0.4}PbI_{2.9}Br_{0.1}$) exhibited a much longer lifespan than $MAPbI_3$ under light irradiation even in 100% oxygen ambience or humid air.

# Introduction

Organic-inorganic hybrid perovskite materials have lately drawn global attention due to their excellent photovoltaic performance[1-7]. The organic cations of the material play an important role in photovoltaic properties such as bandgap, exciton binding energy, and carrier lifetime[8-11], but they also pose an obstacle in the commercialization of perovskite-based solar cells because of their instability. In general, organic compounds can undergo a chemical reaction with oxygen and atmosphere water, which can be activated under certain conditions such as high temperature or light irradiation. Indeed, the hybrid perovskite materials have shown fast decomposition into organic and inorganic components in the presence of water and oxygen under light irradiation[12-20], which casts serious but reasonable doubts to the prospect of using such perovskite materials as an active layer in photoelectric devices.

There have been many studies that attempted improving the chemical stability of perovskite materials, mostly by identifying a single source of instability such as water[16-18,21], oxygen[13-15,19], light[17,20,22], iodine vapor[23] and electric field[17,24]. Recently, we reported a new degradation mechanism[12] based on the electrostatic charges trapped in the perovskite crystal. These charges that were originally generated by light absorption and then trapped mostly along the grain boundaries, were shown to trigger the irreversible degradation of perovskite material in the presence of moisture, which was attributed to the charge driven deprotonation of organic cations.[12] While no similar charge-driven mechanism has been proposed for oxygen-induced degradation of perovskite, we note that the recent discovery of $MAPbI_3$ perovskite suffering from irreversible degradation in the presence of only oxygen without moisture under light soaking[13-15,19] indicates that oxygen and water may individually possess different chemical pathways for perovskite degradation. Although formation of superoxide ($O_2^-$) from the reaction of oxygen with photo-generated electrons was suggested to lead to the deprotonation of organic cations[13-15], the product of deprotonation, methylamine gas, has not been detected, leaving the degradation mechanism in the presence of oxygen and moisture not fully established yet.

In this paper, we first experimentally demonstrate that trapped charges also induce irreversible degradation of $MAPbI_3$ even in the absence of moisture as long as oxygen is supplied. This result and our previous finding about the trapped charge driven degradation in the presence of moisture[12] suggest that the trapped charges lead to the degradation of perovskite material when there is oxygen or water or both. In our experiments, charges were generated either by light soaking or corona ion deposition[12,25], and trapped charges on the grain boundaries were confirmed by Kelvin Probe Force Microscopy (KPFM) measurements for either case of light soaking or ion deposition. To investigate the details of charge-driven degradation mechanism at the atomic level, we carried out AIMD simulations and tracked how oxygen and water molecules in the presence of trapped charges interact with the compositional units of MAPbI$_3$ crystal in the first 2 ps. From these calculations, we confirmed that deprotonation of organic cations could be induced by trapped charges in the presence of water as was suggested in our previous report.[12] We also learned that oxygen molecules greatly weaken the MAPbI$_3$ crystal bonding when there are trapped charges even without water. Furthermore, we found that the structurally more stable multi-component perovskite MA$_{0.6}$FA$_{0.4}$PbI$_{2.9}$Br$_{0.1}$ that was previously demonstrated to be more resistant to moisture is also highly resistant to oxygen. The MA$_{0.6}$FA$_{0.4}$PbI$_{2.9}$Br$_{0.1}$-based solar cell device showed a much longer lifespan than the MAPbI$_3$-based device in pure oxygen atmosphere.

**Results and Discussion**

We employed three experimental apparatuses for our perovskite degradation studies. Using the first apparatus that has a sealed chamber (light blocked) with a gas inlet and outlet (Fig. 1a), we can study the reaction between a gas and perovskite film without any external perturbation. The ambient gas in the chamber was adjusted by the different combination of inlet gases. The second apparatus (Fig. 1b), already used in our previous work[12], is designed for intentional charge trapping in the perovskite film without light soaking by introducing N$_2^+$ corona ions. It is noted that gas molecules such as O$_2$ or H$_2$O are mixed with N$_2^+$ ions in the deposition chamber after N$_2^+$ corona ions generated in the corona chamber are injected into the deposition chamber. N$_2^+$ corona ions are selectively deposited on the perovskite surface by applied electric field in the deposition chamber. In the third apparatus (Fig. 1c), we can study the degradation of perovskite film placed in a gas-controlled chamber under solar-simulated light to investigate the synergetic effect of light soaking and gas molecules.

First, to reconfirm the effect of water molecules on MAPbI$_3$ degradation in dark condition, pure nitrogen gas was humidified through a water bubbler and injected into the sealed chamber where a MAPbI$_3$ film was placed (at the relative humidity of 40% and flow rate of 2 lpm). There was no change in the absorption spectrum of the MAPbI$_3$ film for 24 hours as shown in Fig. 1d, which means that water molecules alone do not cause irreversible degradation of MAPbI$_3$ perovskites in dark condition, as was reported previously[12,21]. However, Fig. 1e and 1f show that either ion charge deposition (without light) or light soaking causes irreversible degradation of MAPbI$_3$ film in 24 hours under the same humidified nitrogen gas. Both ion deposition and light soaking were shown to cause charge trapping along perovskite grain boundary, where degradation was initiated, and our previous study[12] proposed that trapped charge would be the main cause for irreversible degradation occurred under light soaking and moisture (refer to Ahn *et al.* for detailed reasoning[12]). However, it still remains unclear how water molecules interact with MAPbI$_3$ crystal in the presence of trapped charges.

Oxygen molecules have also been suggested as a cause of irreversible degradation in previous reports[13-15,19]. Haque and coworkers demonstrated that MAPbI$_3$ could be decomposed by light activated oxygen (superoxide $O_2^-$) even without moisture[13-15]. In their study, a degradation test was carried out under light soaking condition in the presence of oxygen. In the present study, we were able to examine the effect of oxygen molecules on the degradation of MAPbI$_3$ with either light soaking or ion charge deposition without light.

First, we found that injection of only dry air (N$_2$+O$_2$) without charge deposition or light irradiation did not cause any degradation (Fig. 1g). On the other hand, the MAPbI$_3$ film underwent complete degradation in 24 hours under dry air with ion charge deposition (without light) (Fig. 1h) or under light soaking (Fig. 1i) as was the same in the presence of moisture (Fig. 1e, f). In light of the fact that there was no degradation in nitrogen ambient gas (Supplementary Fig. 1), oxygen molecules are believed to be responsible for chemically interacting with the compositional units of MAPbI$_3$ crystal, as will be discussed in detail later. We also found that oxygen-induced degradation took place along the grain boundaries where charges are trapped, as shown by SEM images of the fresh and degraded samples (Supplementary Fig. 2). Both ion deposition and light soaking result in charges trapped along the grain boundaries, as shown by 3D overlapped KPFM images (Supplementary Fig. 2). Thus, we may hypothesize that trapped charges along grain boundaries trigger irreversible degradation of MAPbI$_3$ perovskites in the presence of water or oxygen.

Although the easy degradation of hybrid perovskite is a widely known problem, there is no clear theory to date that explains its mechanism from the atomic-scale viewpoint. However, if we adopt the trapped charge-driven degradation mechanism, we can readily apply a computational method to simulate how the perovskite structure evolves following charge injection. In this study, we first carried out density functional theory (DFT) calculations to examine the structure of perovskite crystal when different electrostatic charges are injected to simulate the charge trapping following light induced exciton generation. We then performed AIMD simulations with periodic boundary conditions on our DFT structures for the surface degradation of $MAPbI_3$ under $O_2$ or $H_2O$ molecules. (Detailed method is presented in Method Section). In all simulations, we adopted an initial geometry with 2 rigidly fixed $MAPbI_3$ units in the bottom of the crystal, 2 structurally relaxed $MAPbI_3$ units on top of the 2 bottom units to form the surface, and adsorbate molecules on the surface. We bestowed different net charge (+1, 0, or −1) in the unit cell to simulate the trapped charge resulting from the separation of excitons generated by light.

We first carried out simulations for the pure $MAPbI_3$ surface without any adsorbate as a control case, which resulted in no special outcome regardless of the net charge trapped (Supplementary Fig. 3). This means that trapped charges themselves do not cause any degradation if there are no gas molecules. To investigate the effect of water molecules in the presence of trapped charge, we performed simulations for five water molecules randomly placed near the surface (Fig. 2a). In the neutral case with no trapped charge (Fig. 2b), all water molecules evolved out of the $MAPbI_3$ surface and clustered with themselves in 2 ps although some of them were quite solidly embedded initially. However, when an electrostatic charge was injected to the crystal, notable changes occurred in atomic motions. In the positively charged crystal (Fig. 2c), the five water molecules quite evenly placed initially on the $MAPbI_3$ surface started to cluster around a methylammonium cation ($CH_3NH_3^+$). The $CH_3NH_3^+$ cation ended up effectively solvated by water molecules, indicating the dissolution of $CH_3NH_3^+$ cations into the surface water on the bulk solid with trapped positive charges. Such a dissolution process appears prompted by the weakened bond between $CH_3NH_3^+$ cation and the $PbI_6^-$ octahedron unit due to the new electrostatic force field induced by the excess positive charge. In the negatively charged crystal (Fig. 2d), on the other hand, a totally different water-induced degradation pathway was discovered. In contrast to the positively charged crystal, water molecules now aggregate around an iodide anion ($I^-$) that is fast moving away from its stable position on the surface due to the electrostatic expulsion by the excess negative charge. A subsequent encounter between this hydrated $I^-$ anion with water molecules would result in volatile species such as hydrogen iodide (HI) and methylamine ($CH_3NH_2$), as shown in the 2 ps snapshot of Fig. 2d, from proton transfer between $CH_3NH_3^+$ and $I^-$. The calculated bond length of HI at 2 ps in Fig. 2d is 1.616 Å, which is very similar to the known value of 1.609 Å. It is noted that the generation of these volatile compounds in the presence of water molecules is predicted for the first time by simulation although the release of such volatile compounds from the perovskite solid has been previously suggested[8,26,27]. This simulation result for the deprotonation of organic cation initiated by trapped charge is in agreement with the scheme suggested by our earlier work[12], which consists of the following three steps as shown in Supplementary Fig. 4.

$I^- + H_2O \rightarrow HI + OH^-$  (shown at 1.75 ps of Supplementary Fig. 4)

$OH^- + H_2O \rightarrow H_2O + OH^-$  (shown at 1.80 ps and 1.85 ps of Supplementary Fig. 4)

$CH_3NH_3^+ + OH^- \rightarrow CH_3NH_2 + H_2O$  (shown at 1.90 ps and 1.95 ps of Supplementary Fig. 4)

The overall reaction is seen to initiate from proton transfer from one of the surrounding water molecules to an iodide anion (1.75 ps in Supplementary Fig. 4), which is followed by sequential water-to-water proton exchange until it eventually ends up in proton transfer from $CH_3NH_3^+$ to $OH^-$ (1.80 ps ~ 1.95 ps in Supplementary Fig. 4). Such a sequential proton transfer pathway akin to the Grotthuss mechanism through a series of hydrogen bonds may very well be energetically favorable over direct proton transfer from $CH_3NH_3^+$ to $I^-$ due to the stabilization of intermediate and transition states. Our AIMD simulation clearly demonstrates the crucial role of water in the degradation of $MAPbI_3$ crystal in the presence of trapped charges as it enables both the solvation of component ionic species and the sequential proton transfer.

We also investigated the effect of oxygen as it is newly shown to cause complete, rapid degradation of $MAPbI_3$ film following charge deposition (Fig. 1h) or light soaking (Fig. 1i). Although Haque and coworkers[13-15] assumed that defect sites of $MAPbI_3$ crystals provide a trap site for oxygen molecules on their way to superoxide formation, we considered no defect sites in our simulation and examined only surface degradation pathways involving oxygen in the presence of trapped charges. As a test case, we first simulated a $MAPbI_3$ surface with four embedded nitrogen molecules with or without trapped charges (Supplementary Fig. 5). All embedded nitrogen molecules escaped from the surface and flew away within 2 ps regardless of the net charge, which indicates that nitrogen molecules are not involved in any significant interaction with the constituents of MAPbI$_3$ perovskite, as can be easily expected. We then carried out simulations for oxygen molecules with the same geometry as the nitrogen molecules in our test run. In stark contrast to the nitrogen case, most of oxygen molecules persisted near the surface regardless of the net charge (Figs. 3b, 3c, and 3d), which suggests that oxygen is involved in strong chemical interaction with the component species of MAPbI$_3$. Additionally, we observed the formation of oxides in positively (+1) (Fig 3c) and negatively (−1) (Figs. 3d) charged case but not in the neutral case (Fig. 3b). Superoxide species were also observed in the presence of a charge, which is in good agreement with previous studies[15].(See Supplementary Table 1). We had a glimpse of a transient interaction toward oxide formation in the middle of simulation (0.5 ps ~ 1.5 ps in Fig. 3b), but it did not lead to actual bond formation between oxygen and I$^-$ anion or Pb$^+$ cation in the absence of charge. On the other hand, in positively or negatively charged perovskite crystal (Figs. 3c and 3d), the strong interaction of oxygen with the compositional species of MAPbI$_3$ resulted in the formation of stable I−O or Pb−O bonds.

To confirm the generation of oxides, we additionally analyzed the I−O, Pb−O, and O−O distances for atoms of interest (those forming the I−O−O−Pb bond at 0.5 ps of Fig. 3b and 1 ps of Figs. 3c and 3d) throughout the simulation. In the absence of any charge, the I−O and Pb−O distances are quite short (~2.1 Å and ~2.5 Å, respectively) in the early stage of simulation (0 ~ 1 ps) but become longer past 1 ps (Figs. 3e and 3f), whereas the O−O distance, to the contrary, start out as a long bond (~1.4 Å) but becomes shorter past 1 ps (Fig. 3g). These observations mean that oxygen do not form stable bonds with the neighboring Pb or I atoms without charge. On the contrary, in the presence of trapped charge, the I−O, Pb−O, and O−O distances show continued vibrational oscillation, indicating the existence of permanent chemical bonds (Figs. 3e, 3f, and 3g). To determine the superoxide character of the oxygen atoms in question, we carried out Bader population analysis for the 2-ps geometry with a charge of +1, 0, and −1, and confirmed that only oxygen molecules that are bonded to perovskite (oxygen 1 for +1 charged simulation and oxygen 1 and 2 for −1 charged simulation shown in Fig.3) have almost −1 charge (Supplementary Table 1). We additionally examined the vibrational frequencies (or periods) of O−O shown in Fig. 3g in order to verify the formation of oxides. The vibrational period of the O−O bond with no charge is 24.3(±1.0) fs (at 1 ~ 1.5 ps), which is comparable to the experimental value of 21.4 fs (1,556 cm$^{-1}$) corresponding to free gaseous oxygen molecule[28], but it becomes much longer (41.1(±8.8) and 39.7(±6.7) fs, respectively) in positively and negatively charged crystal. These longer vibrational periods or lower vibrational frequencies are indicative of a weaker O−O bond and simultaneously stronger I−O and Pb−O bonds in the presence of charge, which should be associated with trapped-charge driven oxidation.

Overall, our AIMD simulations revealed mechanistic details of oxygen-induced degradation of MAPbI$_3$ crystal with trapped charge, while showing that oxygen molecules alone do not lead to perovskite degradation without trapped charge. From our experimental results and AIMD studies, it is evident that MAPbI$_3$ perovskites even in dry air are no longer resistant to oxygen-induced degradation if they happen to contain trapped charges.

Considering the structural distortion in the MAPbI$_3$ crystal[29-31], the region between the PbI$_6^-$ octahedron units may be the weakest point of infiltration for external gas molecules. Therefore, structural stabilization of perovskite by tuning tolerance factor to yield a cubic phase structure may benefit its chemical resistance against water or oxygen molecules. In view of the improved stability of our structurally more stable multi-component perovskite MA$_{0.6}$FA$_{0.4}$PbI$_{2.9}$Br$_{0.1}$ against water[12] (See Supplementary Fig. 6), we attempted to check its stability against oxygen as well. We prepared the MA$_{0.6}$FA$_{0.4}$PbI$_{2.9}$Br$_{0.1}$ film using our three different experimental apparatuses and measured their absorption spectra and change thereof in 24 hours (Fig. 4a, 4b, and 4c). Surprisingly, these perovskite films showed no sign of degradation under oxygen exposure after 24 hours even with charge deposition and light soaking. To further confirm the improvement in material stability, we measured the photovoltaic conversion efficiency (PCE) of (unencapsulated) devices stored under one sun irradiation in pure oxygen for 36 hours (Fig. 5 and Supplementary Fig. 7), which showed a dramatic improvement over conventional MAPbI$_3$ based device raising the prospect of these multi-component perovskite materials for long-term stability.

**Conclusions**

We investigated the atomistic origin and mechanism for irreversible degradation of hybrid perovskite materials using our new experiments and AIMD simulations. We experimentally demonstrated the validity of trapped charge-driven degradation of perovskite materials in the presence of either water or oxygen by observing changes in the absorption spectra of perovskite film. We also investigated interactions between gas molecules and MAPbI$_3$ crystal in the first 2 ps of perovskite degradation by tracking atomic motions using AIMD simulation. We found that different polarities of trapped charge lead to atomically different routes of degradation. In particular, in the negatively charged crystal, our calculation for water-covered perovskite clearly showed that the deprotonation of $CH_3NH_3^+$ that has so far been only hypothesized in several previous studies can actually occur in water-mediated pathways. On the other hand, in our simulations for oxygen-covered perovskite, we found that trapped charge weakens the Pb–I bond and leads to stronger I–O and Pb–O interactions. Finally, our multi-component structurally stable perovskite material ($MA_{0.6}FA_{0.4}PbI_{2.9}Br_{0.1}$) that had been demonstrated to be highly resistant to moisture was also found to be resistant to oxygen as well. The device employing these perovskites showed far longer lifespans than MAPbI$_3$-based devices even in 100% oxygen environment. Our study suggests that minimizing the trapped charges is crucial, along with designing structurally stable perovskite material, in order to warrant commercially viable perovskite solar cells.

## Method

**Solution preparation.** All solutions were prepared in the nitrogen-filled glove box. For the MAPbI$_3$ solution, equimolar amounts of metyhlammoium iodide (MAI), PbI$_2$, and dimethylsulfoxide (DMSO) (i.e., 159 mg of MAI (Xi'an Polymer Light Technology), 461 mg of PbI$_2$ (Alfa Aesar), and 78 mg of DMSO (Sigma-Aldrich)) were dissolved in 0.6 ml of dimethylformamide (DMF) (Sigma-Aldrich). Our multi-component perovskite solution was prepared in the likewise stoichiometric way, i.e., by dissolving 79.5 mg of MAI, 68.8 mg of formamidinium iodide (FAI), 11.2 mg of MABr, 461 mg of PbI$_2$, and equimolar DMSO in 0.6 ml of DMF. To prepare the hole transport material (HTM) solution, 72.3 mg of Spiro-MeOTAD (Merk), 28.8 μl (26.6mg) of 4-tert-butyl pyridine (Sigma-Aldrich, 96%) and 17.5 μl of lithium bis(trifluoromethanesulfonyl)imide (Li-TFSI) solution (520 mg Li-TSFI (Sigma-Aldrich) in 1 ml acetonitrile (Sigma–Aldrich, 99.8%)) were dissolved in 1 ml of chlorobenzene (Sigma-Aldrich).

**Perovskite film & device fabrication.** Indium tin oxide(ITO)-coated glass (AMG, 9.5Ω cm$^{-2}$, 25×25 mm$^2$) and pure glass (AMG, 25×25 mm$^2$) substrates were sequentially cleaned by sonication in acetone, isopropanol, and deionized water for 15 min each. All perovskite films were fabricated by Lewis base adduct method[32]. To fabricate a perovskite film, the prepared perovskite solution was coated on a pure glass substrate by spin-coating at 4000 rpm for 20 s with diehtyl ether dripping treatment. After spin-coating, the MAPbI$_3$ film was sequentially annealed at 65 °C for 1 min and 100 °C for 4 min. Our multi-component perovskite film was fabricated via the same method and annealed at 130 °C for 20 min. To prepare the electron transport layer (ETL), a 35-nm-thick layer of C$_{60}$ was deposited on the cleaned ITO glass substrate by using a vacuum thermal evaporator (<10$^{-7}$ Torr) at the constant rate of 0.1 Å s$^{-1}$. After the C$_{60}$ deposition, the perovskite layer was coated by the same process. The HTM solution was then coated on the ITO/C$_{60}$/perovskite substrate by spin-coating at 2000 rpm for 30 s. Finally, a 50-nm-thick layer of gold was deposited on the HTM by using the same vacuum thermal evaporator at the constant rate of 0.3 Å s$^{-1}$.

**Experimental setup for degradation test**

The fully sealed rectangular parallelpiped (RP) chamber has an inner dimension of 35 mm width, 35 mm length, and 30 mm height. This chamber is made of stainless steel to prevent gas leakage except for the top with quartz windows that let light pass through. For ion generation and deposition, two connected chambers, each designated IG for ion generation and ID for ion deposition, were used. The IG chamber produces nitrogen ions by corona discharge that uses an asymmetric field to extract electrons from nitrogen molecules. The ions are then transported by the gas flow and deposited on the perovskite film by the electrostatic force exerted by a high negative voltage applied (–2 kV). The IG chamber is cylindrical in shape (30 mm in diameter, 35 mm in height) and made of transparent acrylic material so that corona discharge can be visually monitored. The ID chamber is also cylindrical in shape (50 mm in diameter, 60 mm in height) and made of stainless steel to block all incoming light. The two chambers are connected to the tee tube so that water molecules or oxygen molecules could be introduced from outside.

The rate of injected gas is controlled by a mass flow controller (MKS Instruments, MFC Controller 247D, MFC 1179A). We used highly purified gases of nitrogen (99.999%) and oxygen (99.995%). Dry air consists of 80% nitrogen and 20% oxygen. To inject water molecules, nitrogen gas was passed through a water bubbler and the relative humidity was monitored by a portable multifunction data-logger (Delta OHM, Data logger DO9847, Temp&Humidity probe HP474AC) at the center of the RP chamber and gas exit of ID chamber. The light soaking experiment was carried out under AM 1.5G one sun irradiation by using a 450 W xenon lamp (Oriel Sol3A) which was calibrated by standard Si photovoltaic cell (Rc-1000-TC-KG5-N, VLSI Standards).

**Characterization**

**Absorption spectra measurements.** The absorption spectra of perovskite films were measured by a UV-vis spectrophotometer (Agilent Technologies, Cary 5000) in the wavelength range from 400 to 850 nm.

**J-V curve measurement.** The photocurrent density-voltage characteristics of perovskite solar cells were measured by a solar simulator with 450 W xenon lamp (Oriel Sol3A) and a current meter (Keithley 2400) under AM 1.5G one sun irradiation, in which the light source was calibrated by using standard Si photovoltaic cell (Rc-1000-TC-KG5-N, VLSI Standards). A metal mask with an aperture of 7.29 mm$^2$ was used in the measurements.

**Kelvin probe force microscope measurements.** An atomic force microscope (Park systems, NX10) was used to obtain topography and surface potential images of perovksite films. In the ion-treated sample, nitrogen cations produced by corona discharge were deposited on the perovskite film coated on the ITO substrate for 1 hr. In the case of the light-treated sample, the profile of positive charge accumulation in the perovskite film coated on the ITO/C$_{60}$ substrate was mesured after one sun illumination for 1 h.

**Computational details**

All DFT calculations were performed with the Vienna *ab initio* simulation package (VASP, version 5.3.5),[33] and the results were visualized by the VESTA (visualization for electronic and structural analysis) and the VMD (Visual Melecular Dynamics) program[34,35]. The projector augmented wave (PAW) method[36,37] was used to describe the electron-ion interaction with the kinetic energy cutoff set to 520.0 eV for the plane waves. We used the Perdew-Burke-Ernzerhof (PBE) exchange-correlation functional[38] and weak van der Waals interactions were considered by the zero damping DFT-D3 method of Grimme[39]. We performed spin-unpolarized calculations for the neat, water-, and nitrogen-covered $MAPbI_3$ surface but spin-polarized calculations for the oxygen-covered $MAPbI_3$ surface to consider the triplet-state nature of the ground state oxygen. The Brillouin-zone was sampled with a Γ-centered (4 × 1 × 4) Monkhorst-Pack k-point grid for structural relaxation and only a single Γ-point was used for the AIMD simulation. All AIMD simulations were performed in the canonical ensemble using Nosé thermostat with a temperature of 298 K. The total simulation time of each trajectory was 2 ps with a 2 fs time-step. The charge of oxygen molecule was obtained by summing atomic charges of two oxygen atoms from Bader population analysis.[40]

The reference structure of $MAPbI_3$ unit cell was constructed from the experimentally determined geometry. We reproduced the experimentally verified orthorhombic $MAPbI_3$ unit cell that consists of four $MAPbI_3$ units ($a$ = 8.836 Å, $b$ = 12.580 Å, $c$ = 8.555 Å)[41], and made little modification on lattice parameters to make the tetragonal crystal structure of $MAPbI_3$ at 25 °C ($a$ = $c$ = 8.852 Å, $b$ = 12.444 Å)[41]. The reference unit cell was fully relaxed at first, whose surface was defined by inserting a large vacuum layer (15 ~ 25 Å) along the *b*-axis to prevent the interaction with upper periodic unit cells. Initial geometries for AIMD simulation were prepared by placing 5 $H_2O$, 4 $N_2$, 4 $O_2$, or a null layer at appropriate positions of the $MAPbI_3$ surface (same locations chosen for $N_2$ and $O_2$). After the initial relaxation in the neutral state, the simulation started with a net charge of +1, 0, or −1 in the unit cell by ejecting (+1) or injecting (−1) an electron. In all surface relaxations and simulations, two $MAPbI_3$ units at the bottom of the unit cell were rigidly fixed while all other atoms that compose the upper two $MAPbI_3$ units and the adsorbates were allowed to move.

**Data Availability**

The data that support the findings of this study are available from the corresponding author upon request.

## Acknowledgements

This work was supported by the Global Frontier R&D Program of the Center for Multiscale Energy Systems funded by the National Research Foundation under the Ministry of Science and ICT, Korea (2012M3A6A7054855).

## Author Contributions

K.K., E.L., N.A. S.K.K. and M.C. designed the experiments, and discussed the experimental results. K.K. carried out the device fabrication and stability test. E.L. performed AIMD simulations. J.H. discussed the simulation results. K.B. measured KPFM. The work was led by S.K.K. and M.C., while K.K., E.L., N.A. S.K.K. and M.C. contributed to the writing of the paper.

## Competing financial interest

The authors declare no competing financial interests.

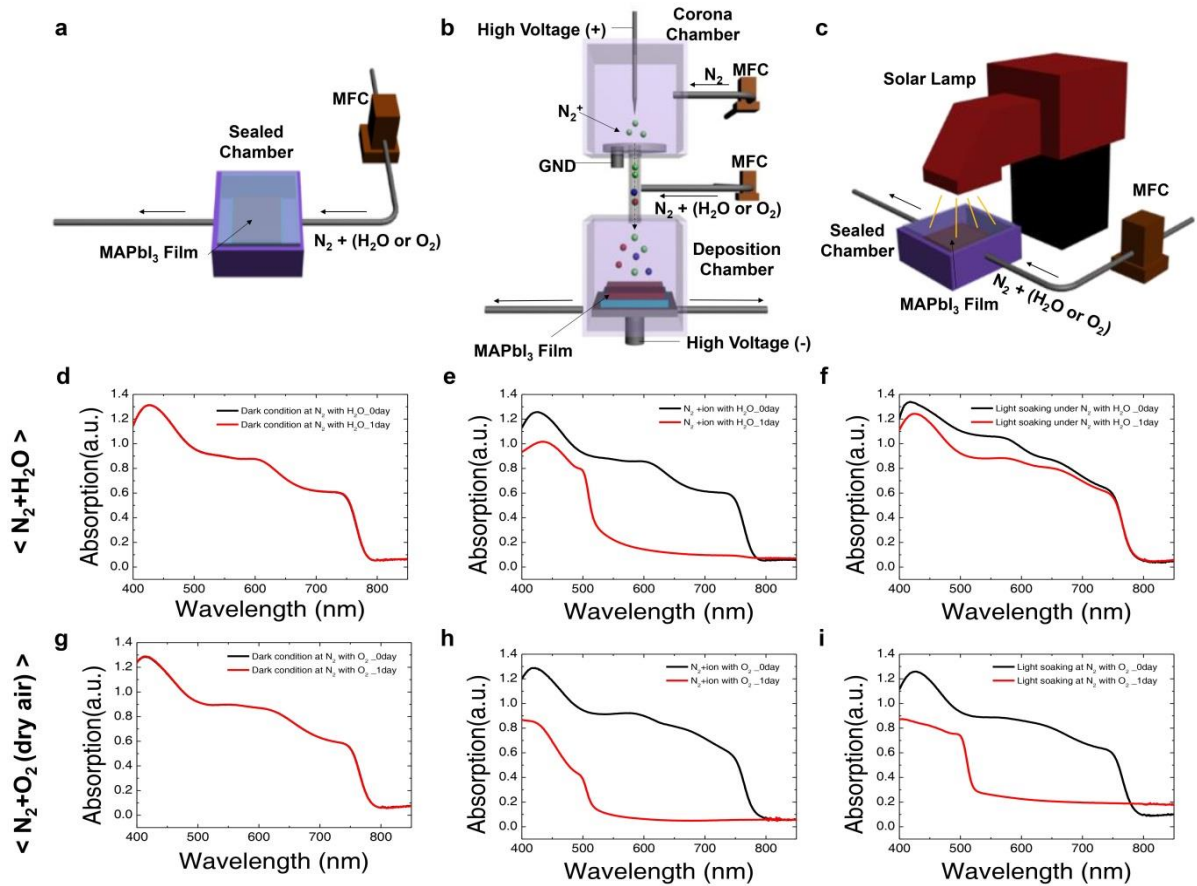

**Figure 1 a-c,** Schematic illustration of experimental apparatuses for aging test, **a,** under dark condition, **b,** with $N_2^+$ deposition, and **c,** under one-sun irradiation. **d-f,** Absorption spectra of MAPbI$_3$ film measured before and after 1 day of aging with a continuous flow of humidified nitrogen gas, **d,** under dark condition, **e,** with nitrogen cation deposition, and **f,** under one-sun irradiation. **g-I,** Absorption spectra of MAPbI$_3$ film measured before and after 1 day of aging with dry air, **g,** under dark condition, **h,** with nitrogen cation deposition, and **i,** under one-sun irradiation.

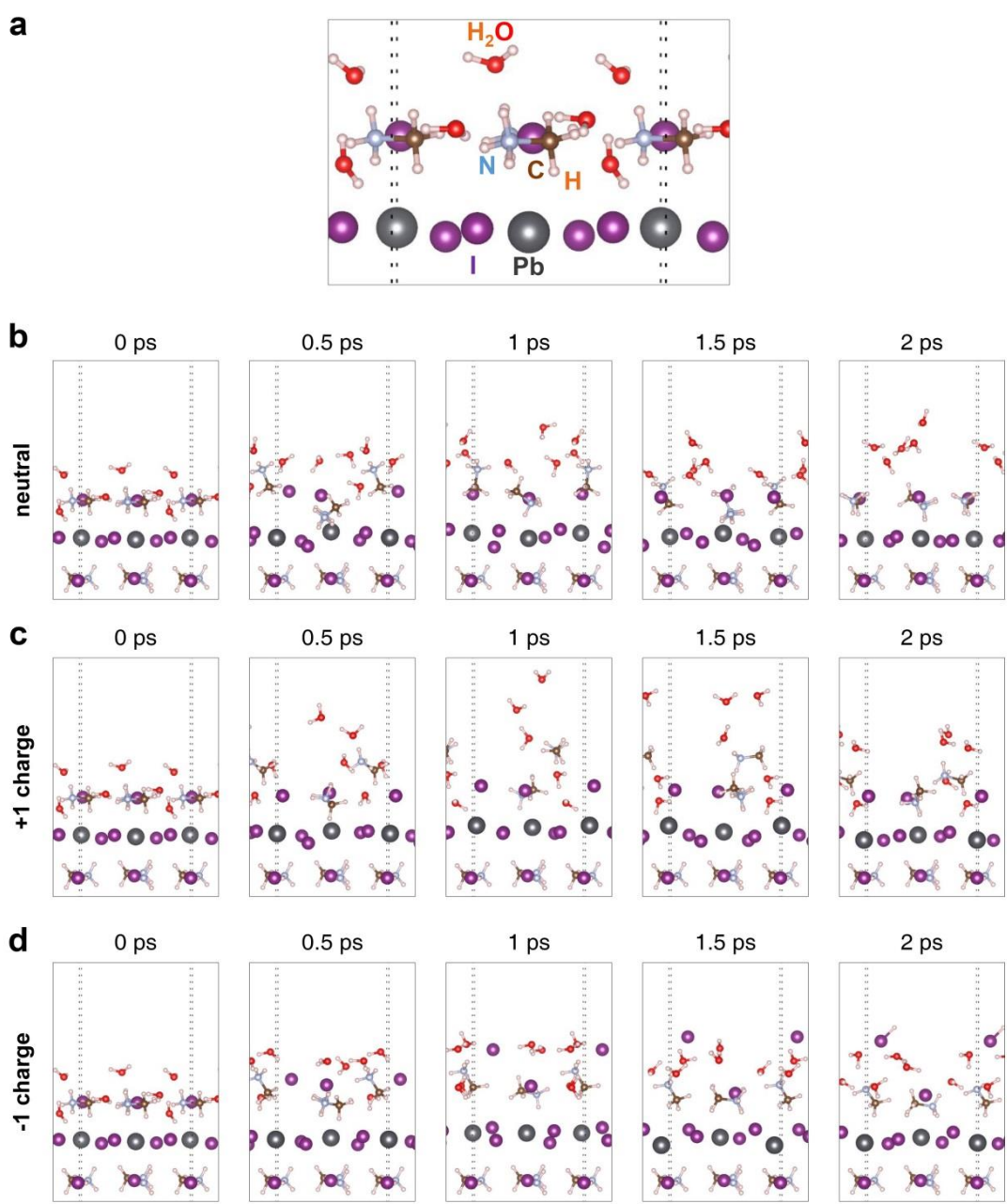

**Figure 2 a,** Expanded view for the initial geometry of 5 H$_2$O molecules on the MAPbI$_3$ surface. **b-d,** Temporal snapshots of the AIMD simulated atomic trajectories of MAPbI$_3$ crystal with a charge of **b,** 0, **c,** +1, and **d,** −1. All simulations start with the same initial geometry at 0 ps shown in **a**. Dotted vertical lines represent the boundaries of actual simulation space, beyond which repeated images of atoms are shown because of the periodic boundary condition.

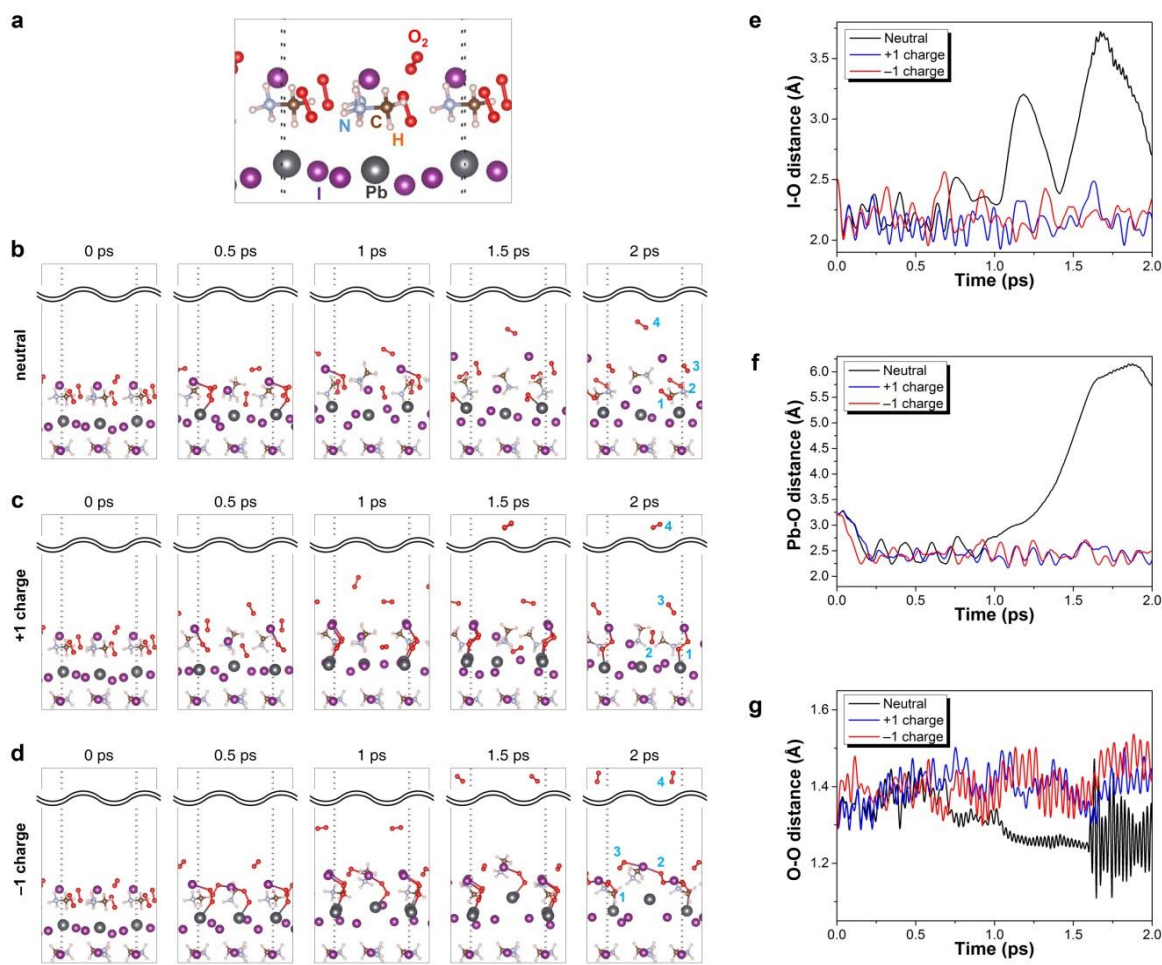

**Figure 3 a,** Expanded view for the initial geometry of 4 $O_2$ molecules on the MAPbI$_3$ surface. **b-d,** Temporal snapshots of the AIMD simulated atomic trajectories of MAPbI$_3$ crystal with a charge of **b,** 0, **c,** +1, and **d,** −1. All simulations start with the same initial geometry at 0 ps shown in **a**. Dotted vertical lines represent the boundaries of actual simulation space, beyond which repeated images of atoms are shown because of the periodic boundary condition. In the 2-ps snapshots, oxygen molecules are designated by numbers for charge analysis (Supplementary Table 1). **e-g,** Temporal change in the **e,** I–O, **f,** Pb–O, and **g,** O–O distance in the I–O–O–Pb bond.

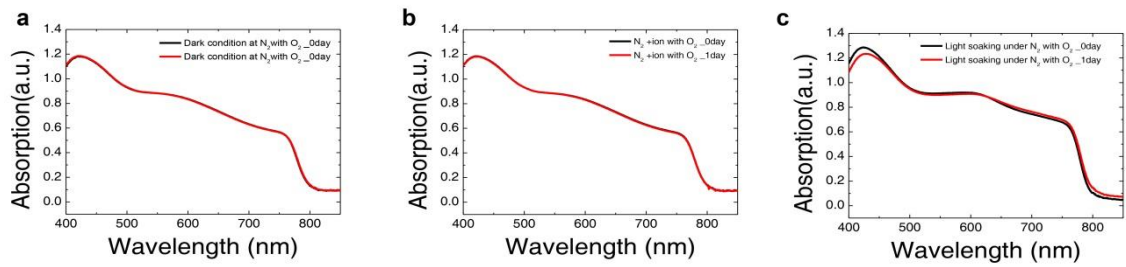

**Figure 4** Absorption spectra of $MA_{0.6}FA_{0.4}PbI_{2.9}Br_{0.1}$ film measured before and after 1 day of aging with dry air, **a,** under dark condition, **b,** with nitrogen cation deposition, and **c,** under one-sun irradiation.

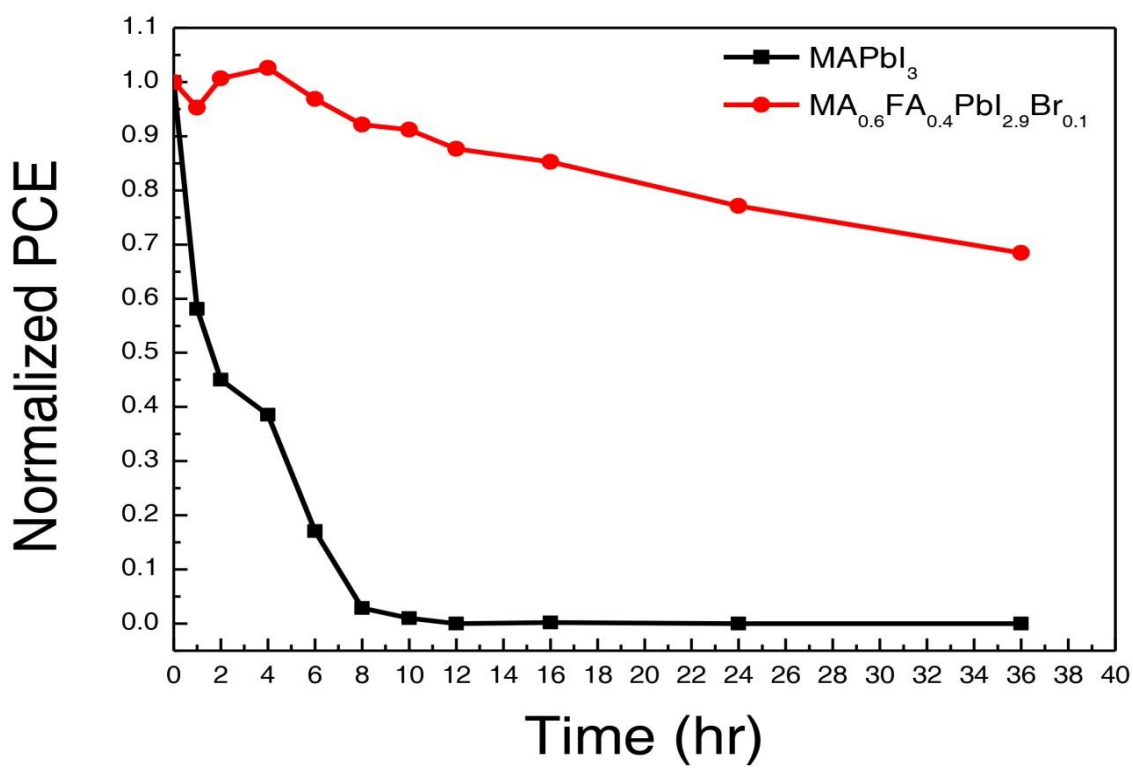

**Figure 5** Normalized PCEs of MAPbI$_3$ and MA$_{0.6}$FA$_{0.4}$PbI$_{2.9}$Br$_{0.1}$ perovskite solar cell under one-sun irradiation with a continuous flow of pure oxygen. In these devices, C$_{60}$ and Spiro-MeOTAD were used for ETL and HTL, respectively.

**Supplementary Information**

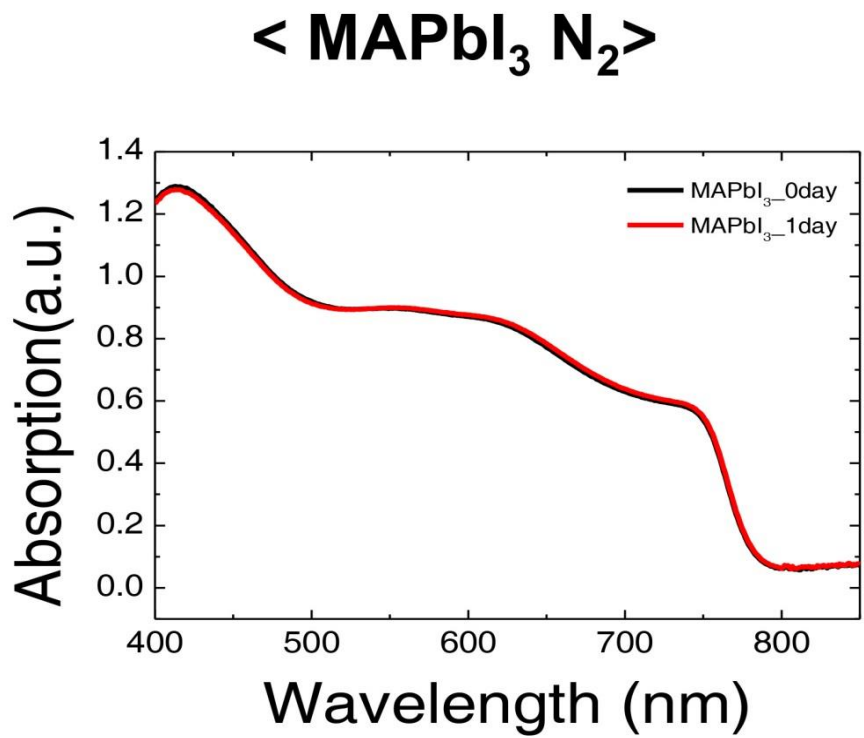

**Figure S1** Absorption spectra of perovskite films aged under one sun irradiation in pure nitrogen ambience.

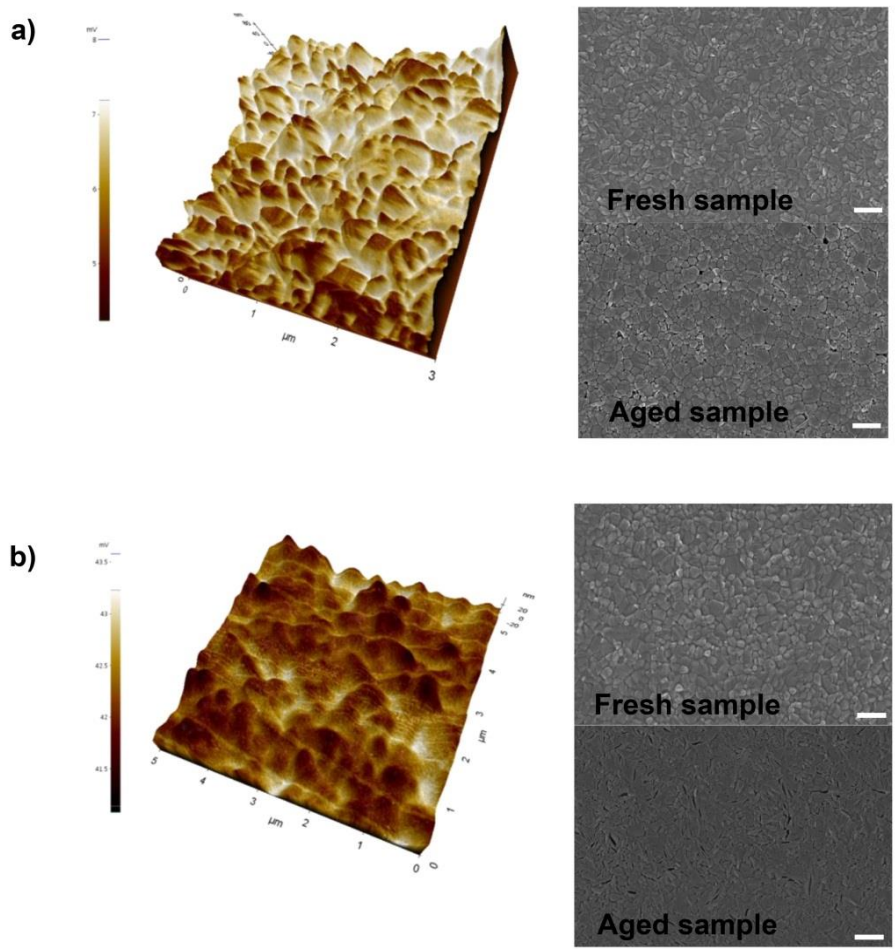

**Figure S2** KPFM and SEM Images of MAPbI$_3$ film after **a,** nitrogen cation deposition and **b,** light soaking

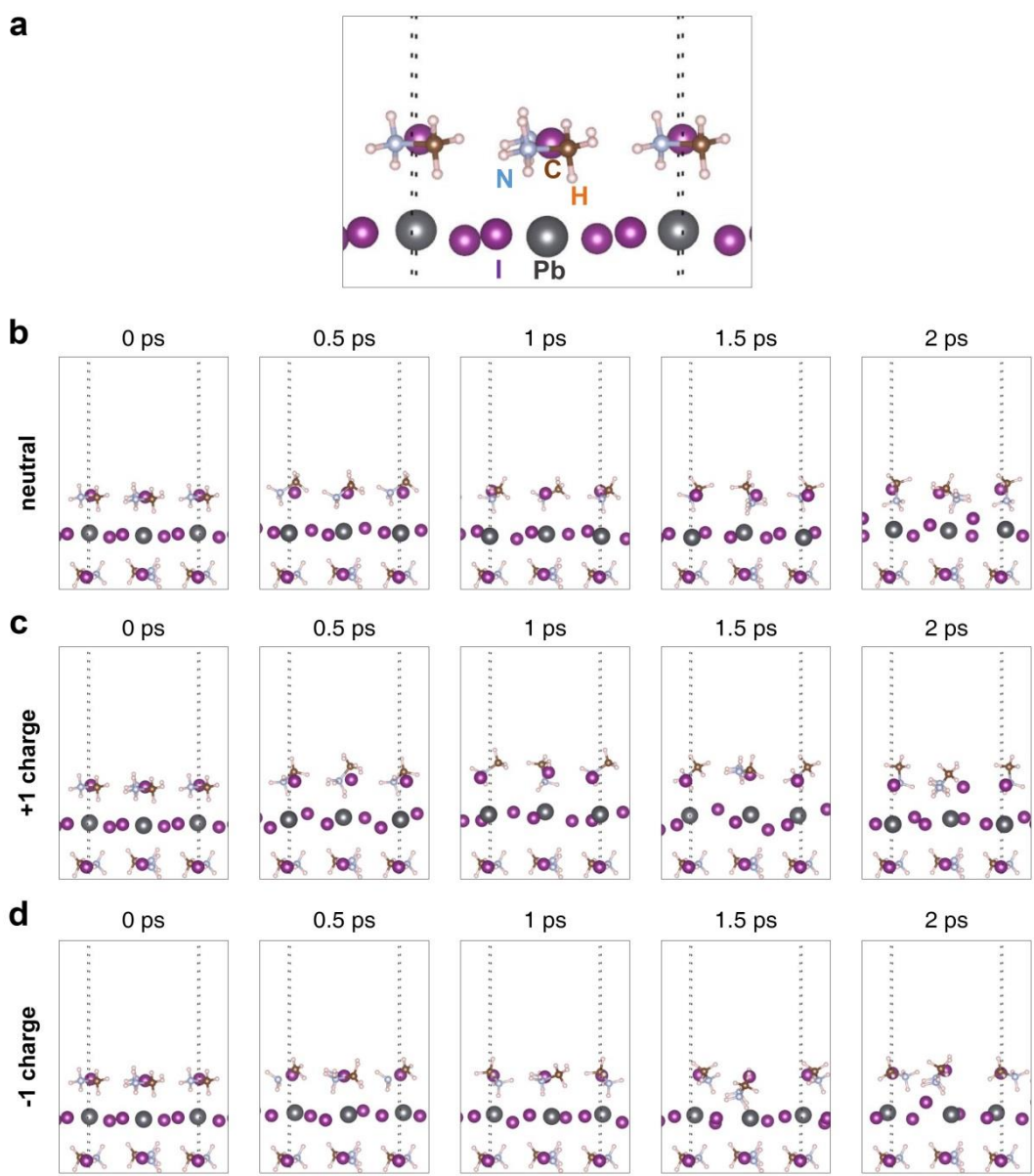

**Figure S3 a,** Expanded view for the initial geometry of pure MAPbI$_3$ surface. **b-d,** Temporal snapshots of the AIMD simulated atomic trajectories of MAPbI$_3$ crystal with a charge of **b,** 0, **c,** +1, and **d,** −1. All simulations start with the same initial geometry at 0 ps shown in **a**. Dotted vertical lines represent the boundaries of actual simulation space, beyond which repeated images of atoms are shown because of the periodic boundary condition.

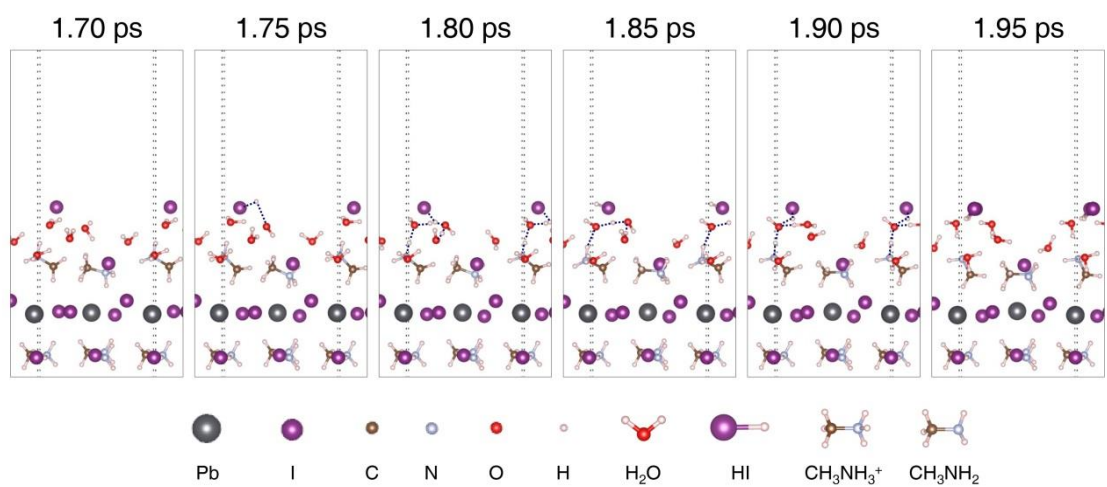

**Figure S4** Temporal snapshots of the AIMD simulated atomic trajectories of 5H$_2$O-covered MAPbI$_3$ crystal with a charge of −1 during the proton transfer. Hydrogen bonds in proximity that form proton wires are shown by navy dotted lines. Dotted vertical lines represent the boundaries of actual simulation space, beyond which repeated images of atoms are shown because of the periodic boundary condition.

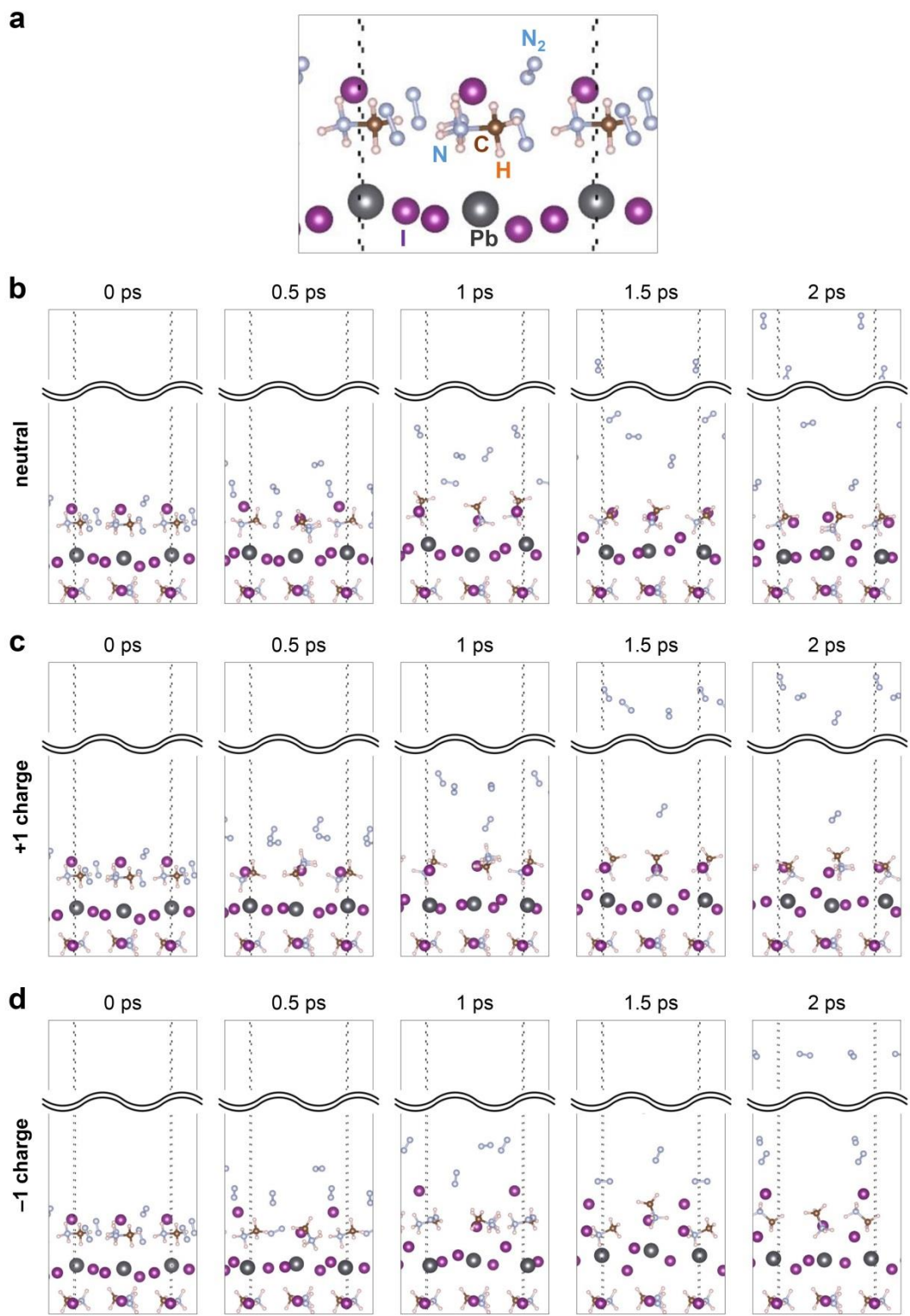

**Figure S5 a,** Expanded view for the initial geometry of 4N$_2$-covered MAPbI$_3$ surface. **b-d,** Temporal snapshots of the AIMD simulated atomic trajectories of 4N$_2$-covered MAPbI$_3$ crystal with a charge of **b,** 0, **c,** +1, and **d,** −1. All simulations start with the same initial geometry at 0 ps shown in **a**. Dotted vertical lines represent the boundaries of actual simulation space, beyond which repeated images of atoms are shown because of the periodic boundary condition.

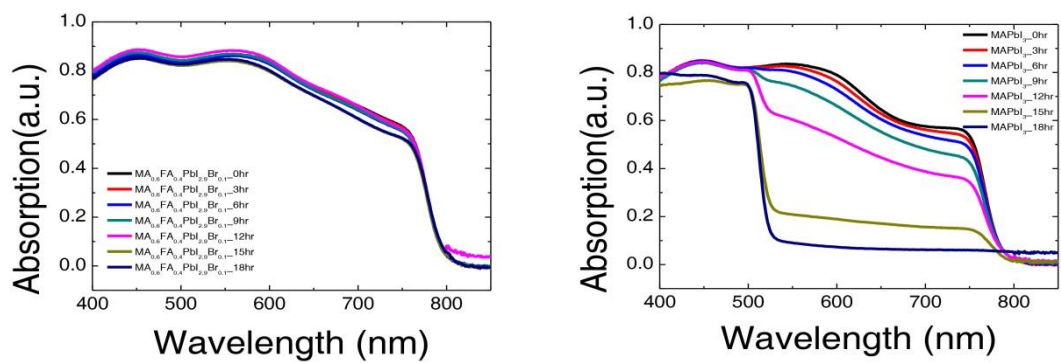

**Figure S6** Time evolution absorption spectra of $MA_{0.6}FA_{0.4}PbI_{2.9}Br_{0.1}$ (left) and $MAPbI_3$ (right) films aged under one sun illumination in humidified dry air for 18 hours.

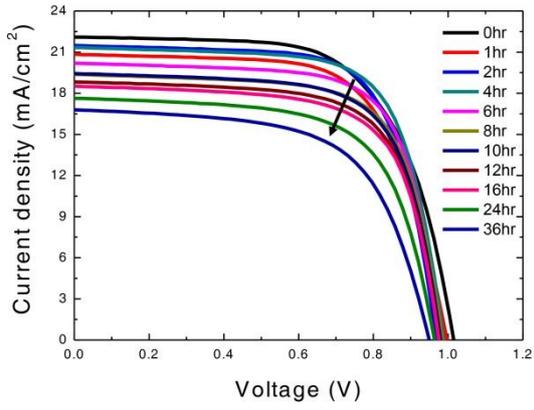 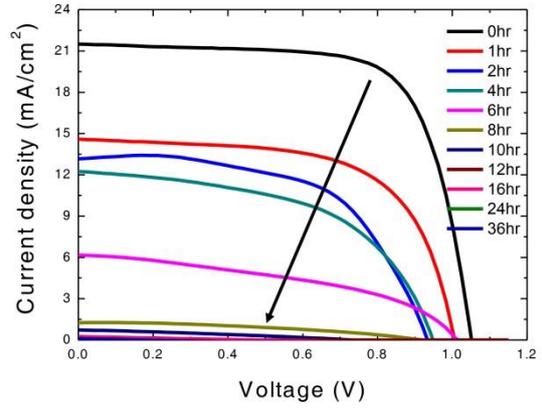

**Figure S7** Time evolution of J-V curves of $MA_{0.6}FA_{0.4}PbI_{2.9}Br_{0.1}$ (left) and $MAPbI_3$ (right) based devices aged under one sun illumination in 100% oxygen ambience for 36 hours.

# Supplementary Table

|  | Net charge |  | Oxygen | | | |
|---|---|---|---|---|---|---|
|  |  |  | 1 | 2 | 3 | 4 |
| 0 |  | status<br>charge | nonbonded<br>−0.42 | nonbonded<br>−0.38 | nonbonded<br>−0.19 | nonbonded<br>−0.02 |
| +1 |  | status<br>charge | bonded<br>−0.94 | nonbonded<br>−0.08 | nonbonded<br>+0.01 | nonbonded<br>+0.19 |
| −1 |  | status<br>charge | bonded<br>−1.04 | bonded<br>−0.98 | nonbonded<br>−0.55 | nonbonded<br>−0.13 |

**Table S1.** Character of chemical interaction with perovskite (top entry) and electrostatic charge (bottom entry in number) of 4 oxygen molecules with the geometry at 2 ps snapshot for $4O_2$-covered $MAPbI_3$ shown in Figure 3.